\documentclass[a4paper,12pt]{article}
\usepackage{fullpage}
\usepackage{graphicx}
\usepackage{amssymb}
\usepackage{amsmath}
\usepackage{multicol}
\usepackage{ulem}
\usepackage{hyperref}

\usepackage{caption}
\usepackage{subcaption}
\newcommand{\fg}{figure }

\newcommand{\pr}{\partial{}}
\newcommand{\bphi}{\mbox{\boldmath $\phi$}}

\newcommand{\be}{\mbox{\boldmath $e$}}

\newcommand{\me}{minimum-energy }
\newcommand{\sk}{Skyrme-Faddeev }
\newcommand{\esk}{extended-Skyrme-Faddeev }
\newcommand{\tc}{topological charge }
\newcommand{\tcs}{topological charges }

\date{}

\title{Static Hopfion solutions of the extended Skyrme-Faddeev model}
\author{David Foster\footnote{email address: dfoster@ursa.ifsc.usp.br}
  \bigskip
  \\Instituto de F\'isica de S\~ao Carlos;
  \\IFSC/USP, Universidade de S\~ao Paulo - USP,
  \\Caixa Postal 369, CEP 13560-970,
  \\ S\~ao Carlos-SP, Brazil}

\begin{document}
\maketitle

\begin{abstract}
We construct non-axially symmetric static soliton solutions, with non-zero topological charges, of an extension of the \sk model. The model has an extra quartic-derivative term and we choose its coupling to the Skyrme-term to be negative. We solve the full equations of motion to find numerical solutions with topological charge up to seven and find that the model favours large ring-like solutions.

\end{abstract}

\section{Introduction}
The Skyrme-Faddeev model \cite{Faddeev1975} is a non-linear model in three-dimensional space where the field takes its value on the unit two-sphere and the \me soliton solutions are called Hopfions. Substantial numerical work, \cite{Sutcliffe:2007ui,Hietarinta:2000ci} etc, has shown that the model has vortex-string like soliton solutions and can resemble either rings, links or knots. Using the Cho-Faddeev-Niemi-Shabanov decomposition, Faddeev and Niemi conjectured that the \sk model describes the low-energy limit of the pure $SU(2)$ Yang-Mills theory \cite{Faddeev:1998eq}. This conjecture is somewhat controversial and there is evidence either way, also Faddeev has recently proposed a modification \cite{Faddeev:2008qe}. \\
When Gies \cite{Gies:2001hk} calculated the Willsonian effective-action, to one-loop, for the $SU(2)$ Yang-Mills theory he discovered the \sk model with an extra quartic-derivative term. This model is known as the \esk model \eqref{extended-SK}, and is believed to have a stronger basis in high-energy physics than the \sk model. \\
The \esk model was first investigated in \cite{Gladikowski:1996mb}, where using an axially symmetric ansatz they found a number of non-trivial Hopfions. This was for a different coupling region to the one investigated here. Later there was another study of the \esk model \cite{Ferreira:2009gj}, here they noticed a different region of stability and using an axially symmetric ansatz they found a number of non-trivial solutions. In \cite{Ferreira:2009gj} the authors found a noticeably different \tc four solution than the one found in the \sk model. Hence this model requires further study. \\
This paper is separated into three sections. In section \ref{The ex model} we shall introduce the \esk model and discuss the two coupling constant regions, in section \ref{knot ansatz} we briefly introduce a technique \cite{Sutcliffe:2007ui} which is used to create topologically non-trivial knotted/linked initial configurations, in section \ref{Numerical solutions} we discuss our charge specific numerical solutions for two coupling regimes. Then there are some concluding remarks.

\section{The \esk model} \label{The ex model}
Here we are only interested in static-\me solutions, so the model is best defined by the static-energy functional
\begin{equation}
 E=\int_{\mathbb{R}^3} \left(\pr_i \bphi \cdot \pr_i \bphi+\frac{\kappa}{2}F_{ij}F^{ij} -\frac{\beta}{2}(\pr_i \bphi \cdot \pr_i \bphi)^2  \right) d^3x, \label{extended-SK}
\end{equation}
where $F_{ij}=\epsilon_{abc}\phi_a \pr_i\phi^b \pr_j \phi^c$.
The field is a three-component unit vector $\bphi=(\phi_1,\phi_2,\phi_3)$ and is the map $\bphi:\mathbb{R}^3 \to S^2$. Finite energy requires that the field tends to a constant at spatial infinity, and we choose this to be $\bphi^\infty=(0,0,1)=\be_3$. This condition compactifies space to $S^3$, hence the field can be extended to the map $\bphi:S^3 \to S^2$. Such maps belong to an element of $\pi_3(S^2)=\mathbb{Z}$, which are indexed by an integer $Q$. This integer is referred to as the topological charge. The \tc can be found in one of two ways. Firstly, the tensor $F_{ij}$ defines a 2-form on the 3-sphere, which because $H^2(S^3)=0$ is a closed and exact form. Therefore we can define a 1-form, $A$, such that $F_{ij}=\pr_j A_k -\pr_k A_j$ and then $Q$ can be found as the integral
\begin{equation}
 Q=\frac{1}{32 \pi^2} \int_{S^3}\varepsilon_{ijk} A_i F_{jk}. \label{topological-charge}
\end{equation}
Generically preimages of points on $S^2$ will be unions of disjoint loops in $\mathbb{R}^3$. It has been shown \cite{BT1982} that $Q$ is equal to the linking number of the loops created as the preimages of two distinct points. 
\\
It can be easily seen that \eqref{extended-SK} is positive definite for $\kappa>0$ and $\beta <0$, this is the coupling region mentioned earlier and studied in \cite{Gies:2001hk}. It has also been shown \cite{Ferreira:2009gj} that \eqref{extended-SK} is positive definite for $\kappa<0, \beta<0$ with the constraint that $2\beta/\kappa \geq 1$. To see this we first re-write \eqref{extended-SK} as
\begin{equation}
 E=\int_{\mathbb{R}^3} \left\{ \pr_i \bphi \cdot \pr_i \bphi-\frac{\kappa}{2}\left(-[F_{ij}F^{ij}-\frac{1}{2}(\pr_i \bphi\cdot \pr_i \bphi)^2] +\frac{1}{2}(\frac{2\beta}{\kappa}-1)(\pr_i \bphi \cdot \pr_i \bphi)^2 \right) \right\} d^3x, \label{positive-energy}
\end{equation}
and, as in \cite{Ward:1998pj}, we define the $3\times3$ matrix $D^{ab}=g^{ij}(\pr_i \bphi^a)(\pr_j \bphi^b)$. This matrix has two non-zero eigenvalues which we choose to be $\lambda_1$ and $\lambda_2$. Then we can instantly see that $\pr_i \bphi \cdot \pr_i\bphi = \lambda_1 + \lambda_2$ and $F_{ij}F^{ij} = 2\lambda_1 \lambda_2$, and we discover the identity
$$
F_{ij}F^{ij}-\frac{1}{2}(\pr_i \bphi \cdot \pr_i\bphi)^2 =-\frac{1}{2}(\lambda_1-\lambda_2)^2.
$$
Therefore \eqref{positive-energy}  is positive definite if 
$$
\kappa<0,~~ \beta<0,~~ 2\beta/\kappa \geq 1.
$$

This is the coupling constant region of interest in this paper. The interest in this region is because $\kappa$ and $\beta$ have the same sign, which is needed to agree with the effective Lagrangian in \cite{Gies:2001hk}. Also, it has been shown \cite{Ferreira:2008nn} that if one restricts to the sector $2\beta/\kappa =1$ and introduces a constraint the model becomes integrable and exact vortex solutions can be constructed. \\
Similar to the Vakulenko and Kapitanski argument \cite{VK1979} for the \sk model, this model also has a lower energy bound \cite{Ferreira:2009gj}. To find this bound we first use the inequality $2 F_{ij} F^{ij} \leq (\pr_i \bphi \cdot \pr_i \bphi)^2$ \cite{Ward:1998pj} (this can be easily found from the above eigenvalues) and re-write the static-energy functional \eqref{extended-SK} as the inequality
\begin{eqnarray}
E &\geq&  \left((\int d^3 x\sqrt{2 F_{ij}F^{ij}})^{\frac{1}{2}} - \sqrt{|\frac{\kappa}{2}(\frac{2\beta}{\kappa}-1)|}(\int d^3 xF_{ij} F^{ij})^{\frac{1}{2}} \right)^2 \\
&+&2^{5/4}\sqrt{|\frac{\kappa}{2}\left(\frac{2\beta}{\kappa}-1\right)|} \left(\int d^3 x \sqrt{F_{ij}F^{ij}}\right)^{\frac{1}{2}} \left(\int d^3x (F_{ij}F^{ij})\right)^{\frac{1}{2}}.
\end{eqnarray}
Now using the Sobolev-type inequality
\begin{equation}
C \left(\int d^3x \sqrt{F_{ij}^2}\right)  \left(\int d^3x F_{ij}^2\right) \geq 8  \left(\frac{1}{32\pi^2} \int d^3x \varepsilon_{ijk}A_iF_{ij}\right)^{\frac{3}{2}},
\end{equation}
where $C$ is a universal constant and using Ward's argument \cite{Ward:1998pj} that $C=1/64 \sqrt{2}\pi^4$ is optimal gives the bound
\begin{equation}
 E \geq 64\pi^2 \sqrt{|\frac{\kappa}{2}(\frac{2\beta}{\kappa}-1)|} Q^{3/4}. \label{E-bound}
\end{equation}

Choosing a point on $S^2$ as the boundary value defines a unique antipodal point and we define the preimage of this point as the location curve. In this case the location curve is the curve $\phi_3=-1$ and is represented by the blue curve in all the images. The red curve represents the preimage of another point near to $\phi_3 =-1$ and is used to show the linking.
\\
In \cite{Ferreira:2009gj} the authors used a toroidal ansatz to reduce the $(3+1)$-dimensional problem to a euclidean $2$-dimensional problem. This greatly reduced the computational power needed to find \me soliton solutions. They proposed a number of \me solutions for $\beta e^2 =1.1$, but because of the ansatz these solutions where constrained to be planar. 
\\
Our research is to build on this by finding \me solutions in full $3$-dimensional space, with no constraint or ansatz. The motivation for this is because it is well known that most of the \tc specific \me solutions of the \sk model are not planar. In fact only the \tcs one, two and four are planar. Hence we expected to identify lower energy solutions for the \tc three and find new solutions for \tcs greater than four. Also, in the usual-\sk model the \me \tc four can be understood as two \tc two Hopfions on top of each other \cite{Hietarinta:2000ci}, but in this model it has been found that the \me solution is a single \tc four ring. Hence we expect to find new soliton solutions for larger topological charges. 
\\

\section{Knotted ansatz}\label{knot ansatz}
To confidently identify \me solutions, we are forced to try a number of initial configurations with the correct topology. We can generate configurations with the correct topology using a technique described in \cite{Sutcliffe:2007ui}. We shall briefly outline this technique here. Firstly, we identify space with $S^3$ via the degree-one spherically-equivariant map,
\begin{equation}
 (Z_1,Z_0)=\left((x_1+ix_2)\frac{\sin f}{r},\cos f +i \frac{\sin f}{r} x_3\right),
\end{equation}
where $f(r)$ is a monotonically decreasing profile function, with the boundary conditions $f(0)=\pi,f(\infty)=0$.\\
This identifies the space as an $S^3\subset\mathbb{C}^2$. We can then identify the stereographic projection, $W$, of $\bphi$ with the rational-map, $p(Z_0,Z_1)/q(Z_0,Z_1)$, of two polynomials $p,q$.  
\begin{equation}
 W=\frac{\phi_1+i\phi_2}{1+\phi_3}=\frac{p(Z_0,Z_1)}{q(Z_0,Z_1)}. \label{Rational map}
\end{equation}
If we choose the polynomials such that the number of preimages of a point is $Q$ \footnote{This is a definition of topological degree.}, we then have a \tc $Q$ configuration \cite{Sutcliffe:2007ui}. We can choose how the soliton `knots' by choosing $q(z)=0$ to be a curve which describes an appropriate knot. The simplest example is the axially-symmetric planar ring which is found from the rational-map
\begin{equation}
 W= \frac{Z_1^n}{Z_0^m}. \label{ring-rat-map}
\end{equation}
This produces a field configuration where $\phi_3=-1$ on a planar ring and has \tc $Q=nm$. Throughout the text we categorise loop-like field configurations as $A_{n,m}$. \\
We can also create knotted field configurations with the rational map
\begin{equation}
 W= \frac{Z_1^\alpha Z_0^\beta}{Z_1^a+Z_0^b}. \label{knot-rat-map}
\end{equation}
This rational-map has charge $Q=\alpha b + \beta a$ and if $a$ and $b$ are coprime positive integers the location $(\phi_3=-1)$ describes an $(a,b)$-torus knot. We denote such a configuration as $\mathcal{K}_{a,b}$. As well as knotted and ring configurations there are also linked solutions. We denote a configuration of a charge $d$ ring linking with a charge $f$ ring $\gamma$ times as $\mathcal{L}_{d,f}^{\gamma,\gamma}$. It is quite simple to see that such a configuration has charge $Q=2\gamma + d+f$. We can create such field configurations with a reducible rational-map, for example we can create a $Q=5$ linked Hopfion with the map,
\begin{equation}
 W=\frac{Z_1^2}{2(Z_1-Z_0)}+\frac{Z_1}{2Z_0}.
\end{equation}

\section{Numerical solutions}\label{Numerical solutions}
In this section we present the numerical technique used to find \me solutions, and we discuss the \me solutions for two choices of $\kappa$ and $\beta$. To find \me solutions of the \esk model we performed a numerical full-field minimisation. This was done using a lattice of $180^3$ points with lattice spacing $\Delta x=0.08$, this size was found to be large enough to contain the soliton with enough accuracy. All of the initial conditions were created using the appropriate rational-map, where $f(r)$ was manually chosen to minimise the initial energy. This is a significantly faster technique than performing computationally intensive Hopfion collisions.

\subsection{Close to the scale-invariant limit}
\begin{figure}[!htb]
\begin{center}$
\begin{array}{cc}
\includegraphics[height=1.9in]{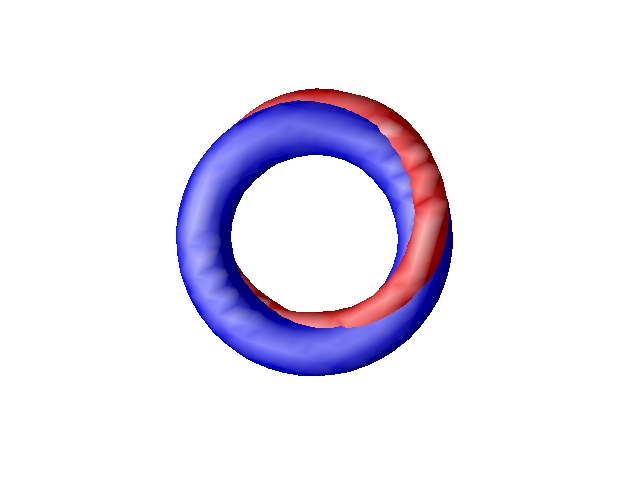} & \includegraphics[height=1.9in]{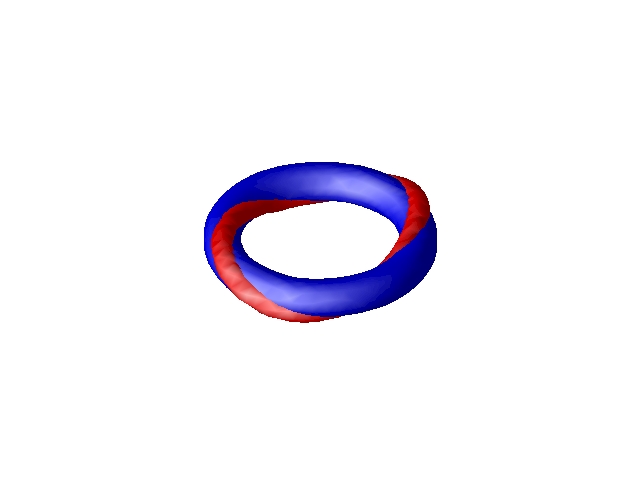} \\
\includegraphics[height=1.9in]{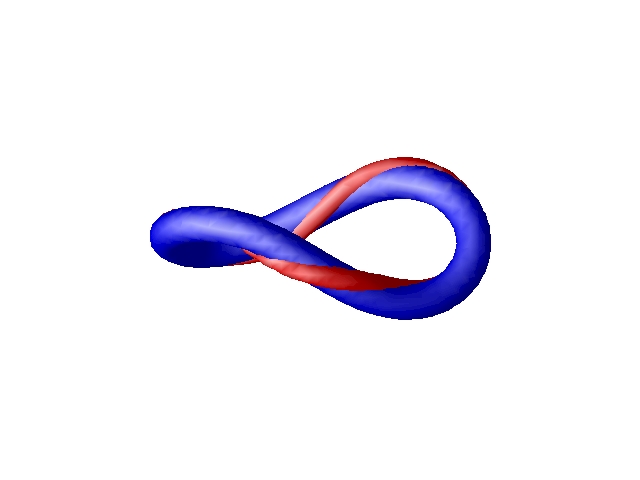} &\includegraphics[height=1.9in]{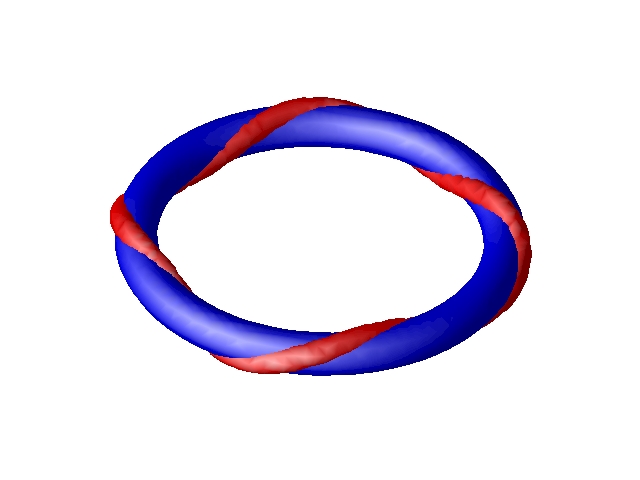} 
\end{array}$ 
\end{center}
\caption{Soliton solutions of the extended-\sk model. The blue surfaces represent the location curve and the red is the second linking curve. Top left is charge one, top right is charge two, bottom left is charge three and bottom right is charge four.}
\label{Extended Skyrme Faddeev solitons}
\end{figure}

In this section we chose $\kappa=-2,\beta=-1.1$, this allows us to directly compare our results with those in \cite{Ferreira:2009gj}. Our numerical solutions are presented in \fg \ref{Extended Skyrme Faddeev solitons} and the energies are shown in table \ref{Table of Energies}.

\begin{table}[!ht]
  \begin{center}
    \begin{tabular}{| l| r |}
    \hline
      Charge  & $E$~~~~\\
\hline
 $Q=1$  & 232.482\\
\hline
 $Q=2$  & 387.432\\
\hline
 $Q=3$  & 573.349\\
\hline
 $Q=4$  & 758.276\\
    \hline
\end{tabular}
\end{center}
\caption{$E$ is the \me solutions found using full three-dimensional minimisation.}
\label{Table of Energies}
\end{table}
Figure \ref{Extended Skyrme Faddeev solitons} shows the buckled $Q=3$ that we expected to discover, which could not be discovered in \cite{Ferreira:2009gj} and has lower energy. Also we have found a similar $A_{4,1}$ \me solution to the one found in \cite{Ferreira:2009gj}, which is different to the $A_{2,2}$ solution found in the \sk model \cite{Sutcliffe:2007ui,Hietarinta:2000ci}. The \me solutions for $Q=1,2,4$ slightly differ the the axial-ansatz solutions in  \cite{Ferreira:2009gj}, the difference might be attributed to numerical accuracy and the extra degrees of freedom of our three-dimensional minimisations.

\subsection{Away from the scale-invariant limit}
A feature of knotted Hopfions is that as the \tc increases the space of saddle point energy solutions increases. This makes finding the \me solutions more involved. An important issue is that as the charge increases there are non-planar knotted solutions which self intersect. It was found that as the soliton tubes dynamically intersected they became unstable for $\kappa=-2,\beta=-1.1$. The actual cause for this instability is not yet understood, but it is believed to be related to the fact that at $\kappa=-2,\beta=-1$ the model becomes integral and has scale-invariance. Potentially the numerical grid is not exact enough to be stable close to this limit. Hence to proceed we arbitrarily fixed $\kappa=-2,\beta=-3/2$, this was found to be stable enough for the knots to self intersect during minimisation. This is the value of the coupling constants used in the remainder of this text.

\begin{table}[!ht]
  \begin{center}
    \begin{tabular}{| l| c | c | c |}
    \hline
      Charge & $E$ &$E/E_{\mbox{bound}}$& Initial $\to$ final configuration~~~~\\
\hline
 $Q=1$ & $552.178$ &$1.236$& $A_{1,1} \to A_{1,1}$\\
\hline
 $Q=2$ & $903.544$ &$1.203$& $A_{2,1} \to A_{2,1}$\\
\hline
 $Q=3$ & $1267.24$ &$1.245$& $A_{3,1} \to \hat{A_{3,1}}$\\
\hline
 $Q=4$ & $1606.32$ &$1.272$&$A_{4,1} \to A_{4,1}$ \\

  & $1687.9$ & ~&$L_{1,1}^{1,1},A_{2,2} \to A_{2,2}$ \\

\hline
$Q=5$  & $1958.73$ &$1.130$& $A_{5,1},\mathcal{K}_{3,2},L_{1,2}^{1,1} \to A_{5,1}$ \\
\hline
$Q=6$  & $2263.04$ &$1.322$& $A_{6,1}, \mathcal{K}_{3,2}\to L_{2,2}^{1,1}$ \\
\hline
$Q=7$  &  $2506.66$ &$1.304$& $L_{1,1}^{2,3}  \to L_{1,1}^{2,3} $ \\
       & $2528.81$  &~& $\mathcal{K}_{3,2} \to \mathcal{K}_{3,2}$ \\
       & $2647.1$  &~& $A_{7,1} \to A_{7,1}$ \\
\hline
\end{tabular}
\end{center}
\caption{The \me found using full three-dimensional minimisation for $\kappa =-2,\beta=-3/2$ and the initial and final forms.}
\label{Table of Energies all}
\end{table}

\begin{figure}[!ht]
       \centering
       \begin{subfigure}[b]{0.3\textwidth}
               \centering
               \includegraphics[width=\textwidth]{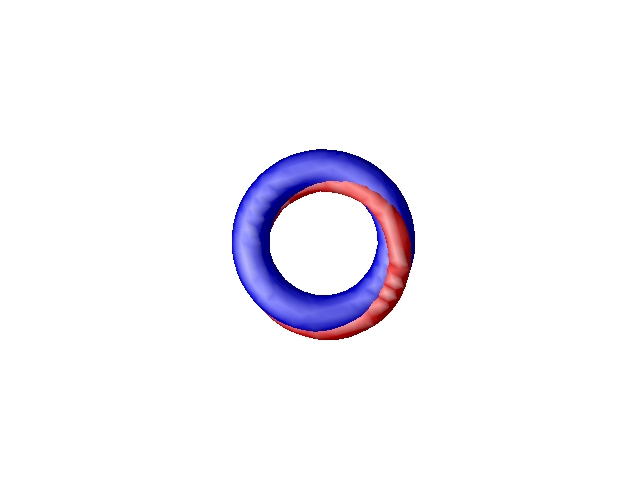}
               \caption{$Q=1 ~ A_{1,1}$}
               \label{Q1A} 
       \end{subfigure}%
       ~ 
       \begin{subfigure}[b]{0.3\textwidth}
               \centering
               \includegraphics[width=\textwidth]{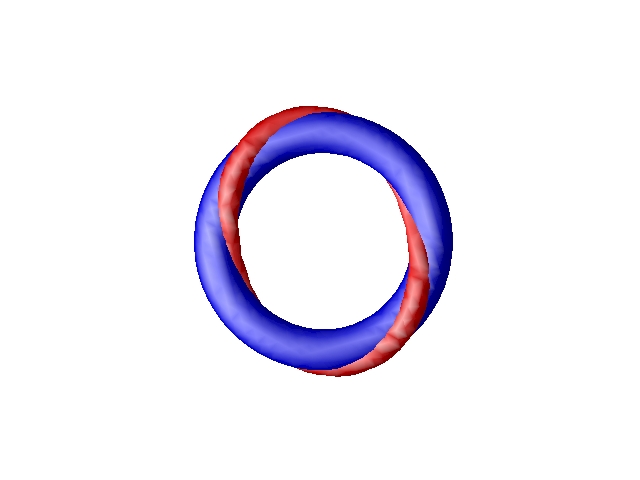}
               \caption{$Q=2 ~A_{2,1}$}
               \label{Q2A}
       \end{subfigure}
       ~ 
       \begin{subfigure}[b]{0.3\textwidth}
               \centering
               \includegraphics[width=\textwidth]{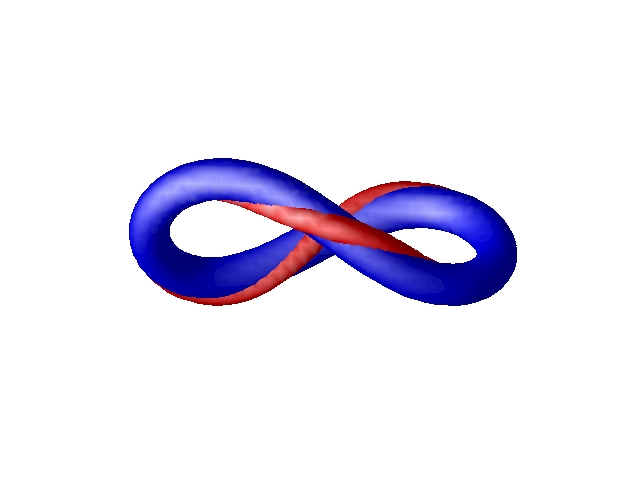}
               \caption{$Q=3 ~A_{3,1}$}
               \label{Q3A}
       \end{subfigure}
       ~ 
       \begin{subfigure}[b]{0.3\textwidth}
               \centering
               \includegraphics[width=\textwidth]{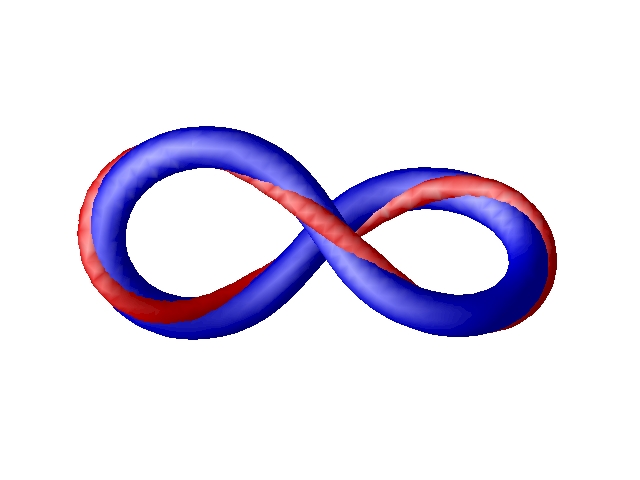}
               \caption{$Q=4~ A_{4,1} \to A_{4,1}$}
               \label{Q4R}
       \end{subfigure}
~ 
       \begin{subfigure}[b]{0.3\textwidth}
               \centering
               \includegraphics[width=\textwidth]{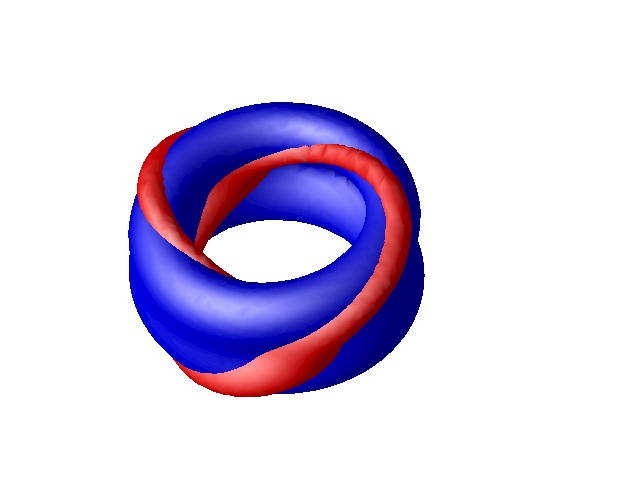}
               \caption{$Q=4~ L_{1,1}^{1,1} \to A_{2,2}$}
               \label{Q4L}
       \end{subfigure}
~ 
       \begin{subfigure}[b]{0.3\textwidth}
               \centering
               \includegraphics[width=\textwidth]{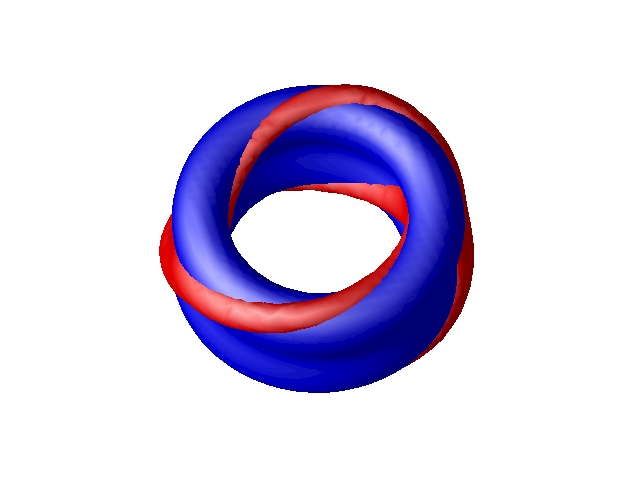}
               \caption{$Q=4 ~A_{2,2} \to A_{2,2}$}
               \label{Q4A}
       \end{subfigure}
~ 
       \begin{subfigure}[b]{0.3\textwidth}
               \centering
               \includegraphics[width=\textwidth]{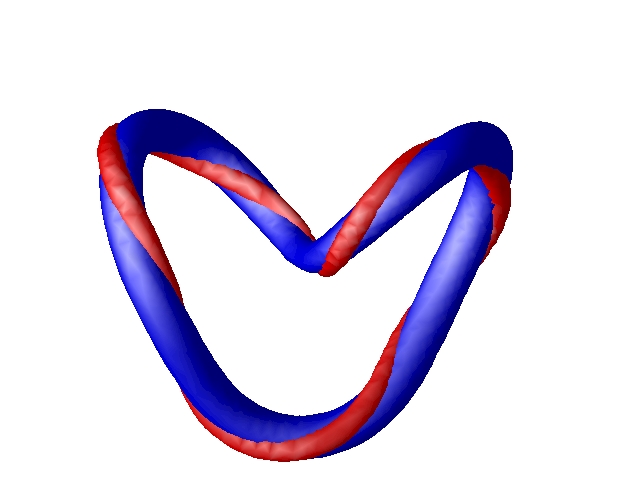}
               \caption{$Q=5 ~K_{3,2} \to A_{5,1}$}
               \label{Q5K}
       \end{subfigure}
~ 
       \begin{subfigure}[b]{0.3\textwidth}
               \centering
               \includegraphics[width=\textwidth]{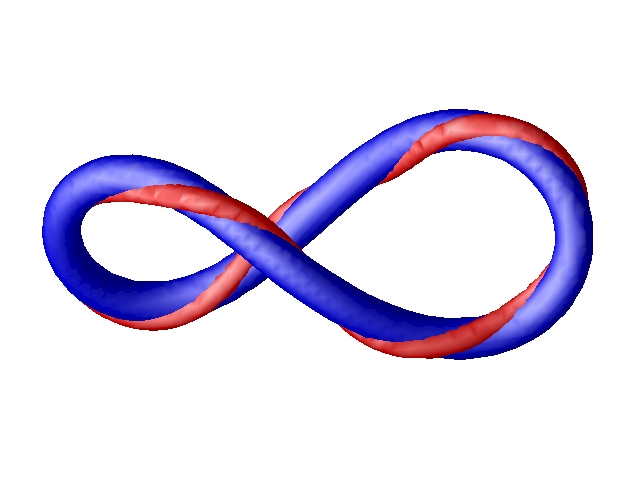}
               \caption{$Q=5 ~L_{1,2}^{1,1} \to A_{5,1}$}
               \label{Q5L}
       \end{subfigure}
~ 
       \begin{subfigure}[b]{0.3\textwidth}
               \centering
               \includegraphics[width=\textwidth]{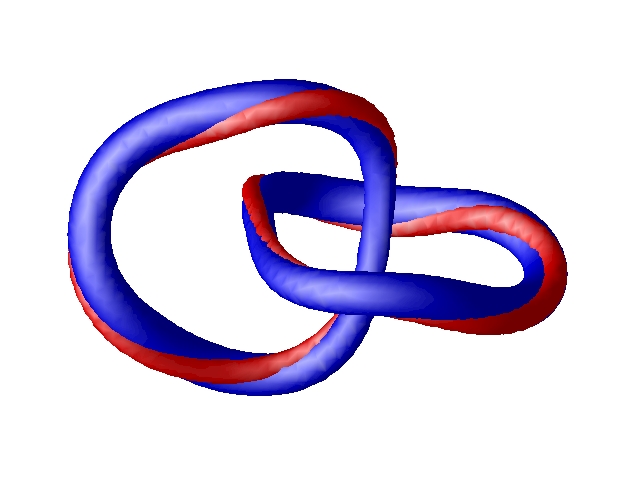}
               \caption{$Q=6 ~ A_{6,1} \to L_{2,2}^{1,1} $}
               \label{Q6}
       \end{subfigure}
~ 
       \begin{subfigure}[b]{0.3\textwidth}
               \centering
               \includegraphics[width=\textwidth]{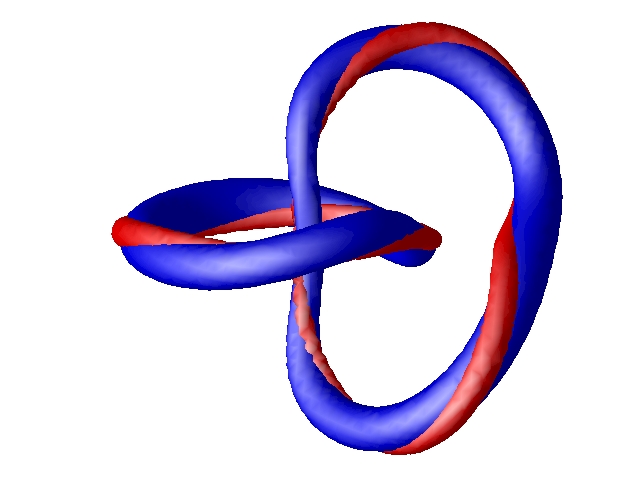}
               \caption{$Q=7 ~ L_{2,3}^{1,1} \to L_{2,3}^{1,1} $}
               \label{Q7L}
       \end{subfigure}
~ 
       \begin{subfigure}[b]{0.3\textwidth}
               \centering
               \includegraphics[width=\textwidth]{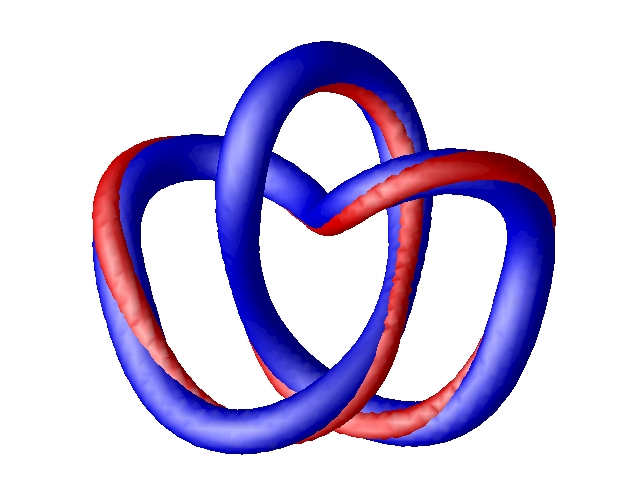}
               \caption{$Q=7 ~ \mathcal{K}_{3,2} \to \mathcal{K}_{3,2} $}
               \label{Q7K}
       \end{subfigure}
~ 
       \begin{subfigure}[b]{0.3\textwidth}
               \centering
               \includegraphics[width=\textwidth]{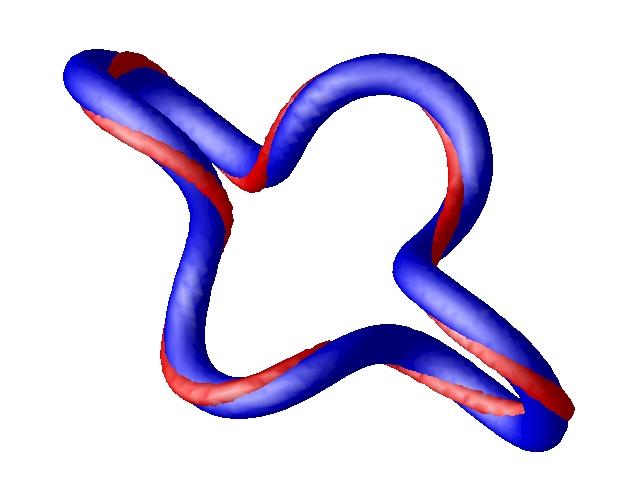}
               \caption{$Q=7 ~ A_{7,1} \to A_{7,1} $}
               \label{Q7}
       \end{subfigure}
    
       \caption{Hopfions of the \esk model for $\kappa=-2,\beta=-3/2$. For each $Q$ the upper most left is the \me solution we found}\label{fig:Hopf}

\end{figure}

The \me solutions we have found for $Q=1,2,3$ are similar to the $Q=1,2,3$ \me solutions of the \sk model. As shown in figure \ref{Q4R} the $Q=4$ \me solution is a larger ring, this is different to the $A_{2,2}$ $Q=4$ \me \sk solution, also for $\beta=-3/2, \kappa =-2$ the ring is buckled. The $\beta=-1.1, \kappa =-2$ $Q=4$ is not very buckled. We have also found linked $Q=4$ solutions which are energetic local minimum or saddle point solutions. From this we can conject that the \esk model energetically prefers large vortex tube configurations and not composite configurations. Again the $Q=5$ \me solution is a large ring (figure \ref{Q5K},\ref{Q5L}), where we have found that a knotted ($K_{3,2}$) and a linked ($L_{1,2}^{1,1}$) initial configurations have deformed to a \me ring-like ($A_{5,1}$) solution. Also, as shown in table \ref{Table of Energies all}, the $Q=5$ \me is much closer to the lower bound than either the $Q=4$ or $Q=6$. The $Q=7$ \me we find is a linked $Q=3$ and $Q=2$ (figure \ref{Q7L}), this solution has $E= 2506.66$ and the $\mathcal{K}_{3,2}$ solution has $E=2528.81$ (figure \ref{Q7K}). These energies are too close to be able to confidently identify the \me solution.

\section{Concluding remarks}
We have presented new \me solution candidates for the \esk model. We have found that for $1\leq Q \leq 7$ the \me solutions are large single-core Hopfion loops. This is notably different from the \me solutions of the \sk model. This difference is not surprising as our coupling is very different to the \sk model, which is known to possess linked and knotted solutions for $Q \leq7$. From the \me solutions found it seems that the \esk model energetically prefers large buckled ring-like field configurations. It has recently been shown that the \me solutions of the \sk model can be approximated by elastic theory \cite{HSS2011}. Here the authors understood the knotting and linking form of \me solutions of the \sk model with a minimal elastic-rod model. They showed that for $Q \geq2$ the \me  solutions are buckled rings. Here we expect the same elastic-rod model to replicate the \me solutions, but with a different coupling parameter and to predict the buckling. This would be very interesting to study.

\section{Acknowledgements}
D.F. would like to thank Prof. L.A. Ferreira for his help and constructive conversation.
\bigskip

\bibliographystyle{utphys}
\bibliography{hopf2.bib}

\end{document}